\documentclass [prb,superscriptaddress, showpacs, twocolumn]{revtex4}
\usepackage{graphicx}
\usepackage{dcolumn}

\makeatletter

\newcommand{\Rmnum}[1]{\expandafter\@slowromancap\romannumeral #1@}
\makeatother

\begin{document}

\title{Coupling of replicate order-parameters in incommensurate multiferroics}

\author{Bruno Mettout}
\affiliation{Laboratory of Physics of Complex Systems, University of Picardie, 33 rue Saint-Leu, 80000 Amiens, France}
\author{Pierre Tol\'{e}dano}
\affiliation{Laboratory of Physics of Complex Systems, University of Picardie, 33 rue Saint-Leu, 80000 Amiens, France}

\date{\today}

\begin{abstract}
The specific properties of incommensurate multiferroic phases resulting from the coupling of order-parameter replicates are worked out using the illustrative example of iron vanadate. The dephasing between the order-parameter copies induces an additional broken symmetry phase corresponding to the lowest symmetry of the system and varies critically at the transition to the multiferroic phase. It reflects the temperature dependence of the angle between paired spins  in the antiferromagnetic spiral structure. Expressing the transition order-parameters in terms of spin-density waves allows showing that isotropic exchange interactions contribute to the stabilization of the ferroelectric phase.
\end{abstract}


\maketitle

\begin{figure}[t]
\includegraphics[scale=0.95]{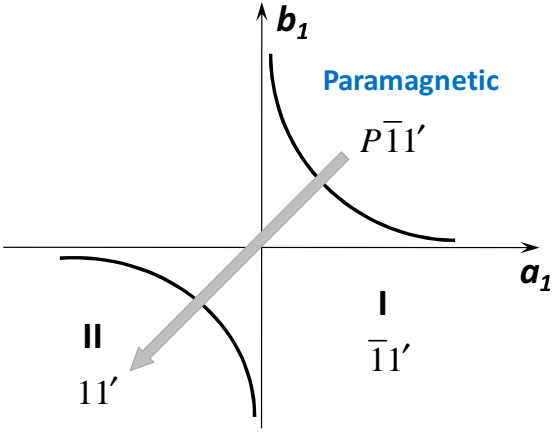}
\caption{(Color online) Phase diagram associated with the free-energy $F$ given by Eg. (\ref{eq:F1}). All curves correspond to second-order phase transitions. The arrow represent the thermodynamic path used for calculating the dephasing $\theta $ given by Eq. (\ref{eq:F2}).}
\label{fig:1}
\end{figure}

\indent One of the distinctive features of magnetic multiferroic materials is the existence of an incommensurate ferroelectric phase induced by the coupling of two antiferromagnetic order-parameters. In multiferroic compounds, such as TbMnO$_3$,~\cite{ref:1} Ni$_3$V$_2$O$_8$~\cite{ref:2} or MnWO$_4$~\cite{ref:3} the two order-parameters display different symmetries, i.e. they transform as distinct irreducible representations (IR's) of the paramagnetic space-group.~\cite{ref:2,ref:4,ref:5} A different situation has been pointed out in the pyroxene NaFeSi$_2$O$_6$,~\cite{ref:6} in which the onset of the ferroelectric phase results from the coupling of two dephased copies of the same order-parameter. The possibility of inducing an additional broken-symmetry phase by replication of a single order-parameter was shown~\cite{ref:6} to reflect the existence of an \textit{effective continuous symmetry} which is broken at the transition to the ferroelectric phase, expressing the continuous rotation of the  phason mode associated with the incommensurate spin-wave order-parameter.\\ 
\indent Phase transitions induced by the coupling of replicate order-parameters display a number of specific features, such as the stabilization of additional broken-symmetry phases and a critical dependence of  the dephasing between the order-parameters.\cite{ref:6} These properties are illustrated here by the theoretical description of the transition to the multiferroic phase observed in iron vanadate FeVO$_4$ which, due to the low triclinic symmetry of the paramagnetic phase,~\cite{ref:7,ref:8} constitutes an unambiguous proof of  replication mechanism. Expressing the transition order-parameter in terms of magnetic spin waves allows showing that the order-parameter dephasing corresponds to the angle between spins forming pairs in the antiferromagnetic spiral structure, and that isotropic exchange interactions contribute to the emergence of the polarization in the multiferroic phase.\\
\indent Starting from the paramagnetic symmetry $P\bar{1}1'$ of FeVO$_4$, a single 2-dimensional IR ($\Gamma$) can be constructed in the general direction $\vec{k}_c=(0.222,0.089,0.012)$ of the triclinic Brillouin-zone reported experimentally,~\cite{ref:7} the matrices of which are given in Table \ref{table:irrep}. A symmetry analysis of the corresponding 2-component order-parameter ($\eta_1 =\rho_1e^{i\theta_1},\eta^*_1=\rho_1e^{-i\theta_1}$) shows that a single second-order transition to an incommensurate non-polar phase of point-group $\bar{1}1'$ (phase $I$) can be induced by $\Gamma $, which coincides with the antiferromagnetic phase observed between $T_P=22K$  and $T_F=15K$ in FeVO$_4$.~\cite{ref:7} Since the propagation vector $\vec{k} $ is nearly temperature-independent, a further symmetry breaking mechanism to the ferroelectric phase stable below $T_F$ cannot be obtained if not assuming that a coupling takes place at $T_F$  between the order-parameter ($\eta_1,\eta^*_1$) and a \textit{dephased} copy ($\eta_2 =\rho _2e^{i\theta_2},\eta^*_2=\rho _2e^{-i\theta_2}$) having the same $\Gamma $-symmetry, as already noted in ref. [8] The free-energy expressing the coupling between the two copies can be written:
\begin{eqnarray}
F = a_1 \rho^2_1 + b_1\rho^2_2 + 2c_1\rho_1\rho_2 cos \theta + \frac{a_2}{2} \rho^4_1 + \nonumber \\
\frac{b_2}{2} \rho^4_2 + \frac{c_2}{2}\rho^2_1\rho^2_2 cos^2 \theta
\label{eq:F1}
\end{eqnarray}
where $\theta=\theta_1-\theta_2$. Fig. \ref{fig:1} shows the phase diagram deduced from the minimization of $F$ with respect to $\rho _1,\rho _2$ and $\theta $. Below the range of stability of phase $I$ ($\theta =n\pi$), the phase diagram contains an additional phase $II$ of magnetic point group $11'$, corresponding to $\theta \neq n\pi $, which displays a spontaneous polarization $\vec{P}$  in general direction, as observed below $T_F$ in FeVO$_4$.~\cite{ref:7,ref:9} The topology of the phase diagram of Fig. 1 is specific to replicate order-parameters, the symmetries of phases $I$ ($\bar{1}1' $) and $II$ ($\bar{1}1'$) representing, respectively, the \textit{minimal} symmetry group induced by $\Gamma $, defined as the invariance group of the general direction in the representation space, and the \textit{lowest symmetry} group induced by $\Gamma + \Gamma $ corresponding to the kernel of the homomorphism of the paramagnetic space group $G_p$ on $\Gamma $.\\ 
\indent The straight thermodynamic path shown in Fig.1 can be parametrized as $a_1=\chi (T-T_F)+4\frac{c^2}{w}$ and $b_1=\zeta (T-T_F)+w$, where the constant parameters $\chi , \zeta $ and $w$ characterizing the path fulfil the condition $\chi \zeta (T_P-T_F)^2+4\frac{c^2_1\zeta }{w}(T_p-T_F)+3c^2_1=0$. It yields the property that the \textit{order-parameter amplitudes} $\rho_1=(-\frac{a_1}{a_2})^{1/2}$ and $\rho_2=(-\frac{b_1}{b_2})^{1/2}$ \textit{vary non-critically} at $T_F$, \textit{while the dephasing $\theta $ plays the role of the critical transition order-parameter}, varying below $T_F$ as:
\begin{equation}
sin\theta = \left (\frac{\chi w}{4c^2} + \frac{\zeta }{w} \right )^{1/2}\sqrt{T_F-T}
\label{eq:F2}
\end{equation}
\begin{table}[t]
\caption{Generators of the irreducible representation $\Gamma $ of the $P\bar{1}1'$ space group for $\vec{k}=(k_x,k_y,k_z)$. $I$ is the space inversion, $T$ is the time reversal and $\vec{R}=(n\vec{a} + p\vec{b} + q\vec{c})$.}
\centering 
\begin{tabular*}{0.48\textwidth}{@{\extracolsep{\fill}} c | c c c} 
$P\bar{1}1'$ & $I$ & $T$ & $\vec{R} $ \\ [0.5ex] 
\hline \\ 
$\Gamma $ & $\left [ \begin{array}{cc} 0 & 1 \\ 1 & 0 \end{array} \right ]$ & $\left [ \begin{array}{cc} -1 & 0 \\ 0 & -1 \end{array} \right ]$ & $\left [ \begin{array}{cc} e^{i\vec{k}\cdot \vec{R}} & 0 \\ 0 & e^{-i\vec{k}\cdot \vec{R}} \end{array} \right ]$ \\ 
\end{tabular*}
\label{table:irrep} 
\end{table}
This property differs from the standard behaviour found in multiferroic phases induced by the coupling of order-parameters having different symmetries where the dephasing $\theta = \pi /2$ is constant and the order-parameter amplitudes $\rho _i$ vary critically with temperature. We show below that the critical variation of $\theta $ can be interpreted as the temperature dependence of the angle between paired spins which are parallel in phase $I$ and become non-parallel in phase $II$.
\indent The equilibrium polarization deduced from the coupling free-energy $F=\alpha_iP^i\rho_1\rho_2 sin\theta+\lambda_{ij}P^iP^j$, gives in triclinic basis:
\begin{equation}
P^j = -\frac{1}{2}(\lambda^{-1})^{ij}\alpha_j\rho_1\rho_2 sin \theta
\label{eq:F3}
\end{equation}
Close to $T_F$ one gets $P^j=\Lambda^j \sqrt{T_F-T}$ with $\Lambda^i=-\frac{1}{2} \left (\chi w+4c^2\frac{\zeta }{w} \right )^{1/2}(\lambda^{-1})^{ij}\alpha_j$. This power law and the correlated Curie-Weiss like sharp peak of the dielectric permittivity $\varepsilon $   are consistent with the dependence on temperature of $\vec{P}(T)$ and $\varepsilon (T)$ observed on single crystal~\cite{ref:7} and polycrystalline~\cite{ref:9} FeVO$_4$ samples. It confirms the hybrid \textit{pseudo-proper} character~\cite{ref:10} of the ferroelectric transition which behaves critically as a proper ferroelectric, although the small value of $\mid \vec{P}\mid $   found below $T_F$ (6-12 $\mu Cm^{-1}$) is of the order of magnitude of an improper ferroelectric. Since inversion symmetry is broken at the transition the system displays two ferroelectric domains~\cite{ref:11} with opposed polarizations corresponding to \textit{opposed dephasings} $\pm \theta$.\\
\indent Application of a magnetic field does not give rise to spectacular magnetoelectric effects in FeVO$_4$, since only trivial couplings between the order-parameter and the magnetic field are allowed by the low symmetry of the system. Thus, biquadratic couplings of the order-parameter to the magnetic free-energy $\mu_{ij}M^iM^j$ yield a renormalization of the $a_i, b_i$, and $c_i$ coefficients in Eq. (\ref{eq:F1}) explaining the observed shift of $T_F$ to lower temperatures~\cite{ref:8} with increasing magnetic field. Standard magnetoelectric couplings $P^iM^jM^k$ are responsible of the progressive reduction and suppression of the multiferroic phase reported at high field.~\cite{ref:12} A different manifestation of magnetoelectricity can be found in the toroidic effects which can be predicted in the ferroelectric phase of FeVO$_4$. Under applied field $\vec{B} $ the toroidal free-energy is:
\begin{equation}
F^T = g_{ij}T^iT^j + \alpha_{ij}T^iB^j\rho_1\rho_2 sin\theta 
\label{eq:F4}
\end{equation}
where $T^l=\kappa ^l_jB^j$ are the induced toroidal moment components and $\kappa^l_j=-\frac{1}{2}\rho _1\rho_2sin\theta (g^{-1})^{li}\alpha_{ij} $ the  \textit{magnetotoroidal} susceptibility components. It yields:
\begin{equation}
T^l = K^l \sqrt{T_F-T} 
\label{eq:F5}
\end{equation}
with $K^l=-\frac{1}{2} \left (\chi w + 4 \frac{c^2\zeta }{w} \right )^{1/2} \left ( g^{-1} \right )^{li}\alpha_{ij}B^j$.  Eq. (\ref{eq:F5}) shows that at constant field the toroidal moment varies critically as the spontaneous polarization at zero fields. Since all components of the magnetotoroidal tensor are non-zero the induced toroidal moment should not orient along $\vec{B} $.\\
\indent To gain insight into the microscopic spin states forming the magnetic structures, one can express the transition order-parameter components in terms of spin-waves. The triclinic centred unit cell (Fig. \ref{fig:2}) in which the inversion centre is assumed to coincide with the centre of the cell, contains 6 magnetic atoms in positions ($1,2,3$) and ($\bar{1},\bar{2},\bar{3}$). Denoting $\vec{s}^m_{npq,i} $ the spin of the $i$ atom in the cell centred at $\vec{R}_{npq} = n\vec{a} + p\vec{b} + q\vec{c}$, with $m=1$ or $m=-1$ for atoms ($1,2,3$) or ($\bar{1},\bar{2},\bar{3}$), respectively, one can write:
\begin{eqnarray}
\vec{s}^m_{npq,i} = \frac{1}{2}\left [\vec{U}_j + im\vec{W}_j \right ] e^{i\vec{k}\cdot \vec{R}_{npq}} + \nonumber \\
\frac{1}{2}\left [\vec{U}^*_j-im\vec{W}^*_j \right ]e^{-i\vec{k}\cdot \vec{R}_{npq}  }     
\label{eq:F6}
\end{eqnarray}
where $\vec{U}_j$ and $\vec{W}_j$ are spin-waves transforming as $\Gamma $ for each $j$ and each space component. It yields 18 spin-waves which are copies of $\eta$ and $\eta^*$. Eq. (\ref{eq:F6}) can be written in real form as: $\vec{s}^m_{npq,i} = \rho^{\mu}_{1,i}cos \left (\vec{k}\cdot \vec{R}_{npq} + \frac{\theta ^{\mu }_{1,i}}{2} \right )\vec{a}_{\mu } - m\rho^{\mu }_{2,i}sin \left ( \vec{k}\cdot \vec{R}_{npq}-\frac{\theta^{\mu }_{2,i}}{2} \right )\vec{a}_{\mu }$, where $\vec{a}_{\mu }=\vec{a},\vec{b},\vec{c}$ for $\mu =1,2,3$ with $U^{\mu}_i=\rho^{\mu }_{1,i}e^{i\theta ^{\mu }_{1,i}}$ and $W^{\mu}_i=\rho^{\mu }_{2,i}e^{i\theta ^{\mu }_{2,i}}$. In phase $I$ all $\vec{U}_j$ and $\vec{W}_j$ are in-phase and one can chose a domain where they are all real. In this domain the inversion centre is located at the centre of the unit-cell ($n=p=q=0$) and the spins take the simpler form:
\begin{figure}[t]
\includegraphics[scale=0.43]{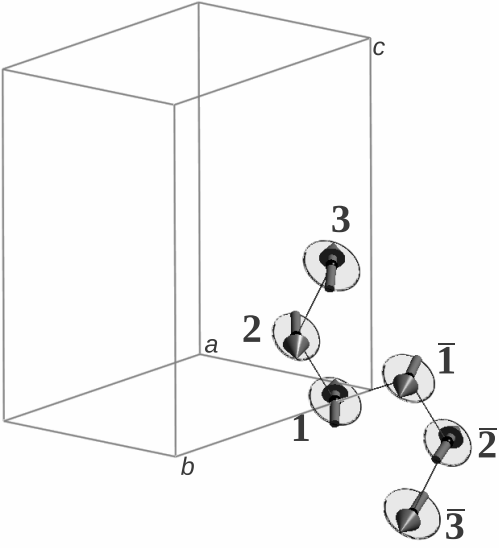}
\caption{Positions of the Fe atoms in the triclinic unit-cell of FeVO$_4$. Atoms forming the pairs $(1,\bar{1}), (2,\bar{2})$ and $(3,\bar{3})$ are related by inversion.}
\label{fig:2}
\end{figure}
\begin{equation}
\vec{s}^m_{npq,i} = \vec{U}_i cos(\vec{k}\cdot \vec{R}_{npq}) - m\vec{W}_i sin(\vec{k}\cdot \vec{R}_{npq})     
\label{eq:F7}
\end{equation}
analogue to Eq. (\ref{eq:F1}) of ref.[7] In order to estimate the dephasing $\theta $ between the order-parameter replicates let us consider the six spins contained in the unit-cell with $n=p=q=0$, namely $s^{\pm}_j$ with $j=1,2,3$. In the non-polar phase $I$ the inversion centre present at the centre of the cell imposes that $s^+_j=s^-_j$ for all $j$, although $s^{\pm}_1\neq s^{\pm}_2\neq s^{\pm}_3$. In the polar phase $II$ the loss of inversion centre yields non-parallel spins for each $s^{\pm }_j$ pair, and one has:
\begin{equation}
s^{+\mu} - s^{-\mu}= 2 \rho^{\mu }_{2,j}sin \theta^{\mu }_{2,j}     
\label{eq:F8}
\end{equation}                                                     
where the angles $\theta^{\mu }_{2,j}$ are replica of $\theta$, all $sin \theta^{\mu }_{2,j}$ varying critically with temperature at the $I\rightarrow  II$  transition as $sin \theta $ (Eq. (\ref{eq:F2})).\\
\indent It has been suggested in ref. (7) that Dzialoshinskii-Moriya (DM) interaction may not be the sole active mechanism at the transition to the multiferroic phase of FeVO$_4$. This property can be verified by considering only one of the two-atoms sublattices, as for example the sublattice involving the $(1,\bar{1})$ pair, and neglect its interactions with the two other sublattices. Since the atoms of a single pair are related by inversion, no DM Interaction can take place. We will now show that a polarization can be induced in the sublattice when assuming only exchange interactions between the two spins. This will demonstrate in turn that purely exchange contributions (one for each sublattice) participate to the total polarization in FeVO$_4$ in addition to the DM and other frustration contributions.\\
\indent Denoting $s^{m,\mu }_{\vec{n} }$ the $\mu $-component of the spin $\vec{s}^m_{\vec{n}}\equiv \vec{s}^m_{npq}(m=\pm 1)$ in the unit-cell $\vec{n}=n\vec{a} + p\vec{b} + q\vec{c}$ the Hamiltonian of the system becomes:
\begin{equation}
H = \sum_{\vec{n},\vec{n}'} \sum_{\mu,\mu 'm,m'} h^{m,m',\mu,\mu '}_{\vec{n}-\vec{n}'}s^{m,\mu}_{\vec{n}}s^{m',\mu'}_{\vec{n}}  
\label{eq:F9}
\end{equation}
where $h^{m,m',\mu,\mu '}_{\vec{n}-\vec{n}'}=h^{-m,-m',\mu,\mu '}_{\vec{n}-\vec{n}'}$ due to inversion symmetry. At low temperature the spin modulus is frozen and only the orientations of the spins can vary. Neglecting quantum effects and higher harmonics the ground state of the system is obtained by minimizing $H$ restricted to the first harmonic with respect to the spin directions. Inserting (\ref{eq:F7}) into (\ref{eq:F9}) one has for $\vec{k}_c \neq 0$:
\begin{equation}
H = \sum_{\mu,\mu',m,m'} (U^{\mu } + imW^{\mu })(U^{\mu'*}-im'W^{\mu'*})A^{\mu\mu'}_{mm'} + cc
\label{eq:F10}
\end{equation}
where  $A^{\mu\mu'}_{m,m'}=\sum_{\vec{n},\vec{n}'} h^{m,m'}_{\vec{n}-\vec{n}',\mu,\mu'}e^{i\vec{k}(\vec{R}_n-\vec{R}'_n)}=A^{\mu\mu'*}_{-m,-m'}$. When only isotropic exchange interactions are involved $h^{m,m',\mu,\mu'}_{\vec{n}-\vec{n}'}=h^{m,m'}_{\vec{n}-\vec{n}'}\delta^{\mu,\mu'}$ and $A^{\mu\mu'}_{m,m'}=A_{m,m'}\delta^{\mu,\mu'}$. Therefore $H$ reduces to:
\begin{equation}
H = \alpha \vec{U}\cdot \vec{U}^* +  \beta \vec{W}\cdot \vec{W}^* + \gamma (\vec{U}\cdot \vec{W}^* + \vec{W}\cdot \vec{U}^*)    
\label{eq:F11}
\end{equation}
where $\alpha=2(A_{11}+A_{1-1}+A_{-11}+A_{-1-1})$, $\beta =2(A_{11}-A_{1-1}-A_{-11}+A_{-1-1})$ and $\gamma=2i(-A_{1-1}+A_{-11})$. Writing the complex vectors as $\vec{U}=\vec{u}_1+i\vec{u}_2$ and $\vec{W}=\vec{w}_1+i\vec{w}_2$, the conditions of constant spin modulus $\parallel s^m_{\vec{n}}\parallel=s_0$ read:
\begin{eqnarray}
u^2_1+w^2_2 = u^2_2+w^2_1=s_0/2,\vec{u}_1\cdot \vec{w}_2 + \vec{u}_2\cdot \vec{w}_1=0 \nonumber\\
\vec{u}_1\cdot \vec{u}_2=\vec{w}_1\cdot \vec{w}_2, \vec{u}_1\cdot \vec{w}_1=\vec{u}_2\cdot \vec{w}_2    
\label{eq:F12}
\end{eqnarray}
and $H$ becomes:
\begin{equation}
H=\alpha(u^2_1+u^2_2) + \beta (w^2_1+w^2_2) + 2\gamma(\vec{u}_1\vec{w}_1 + \vec{u}_2\vec{w}_2)   
\label{eq:F13}
\end{equation}
Phase $I$ corresponds to $\vec{u}_1=\pm \vec{u}_2$ and $\vec{w}_1+\pm \vec{w}_2$. Accordingly, the exchange Hamiltonian has the constant value $H_0=\frac{s_0}{2}(\alpha + \beta)$. The stability of this phase can be tested by varying slightly the four vectors $\vec{u}_1,\vec{u}_2,\vec{w}_1,\vec{w}_2$. Expanding the constraints given by Eqs. (\ref{eq:F12}) to the first-order shows that only five independent parameters ($\omega ^2,\delta u_{1x,y,z},\delta u_{2x,y},\delta w_{1y},\delta w_{2y}$), where $\omega ^2=\delta u^2_{1x}+\delta u^2_{2x}+2\delta u^2_{1z}$, determine the stability of phase $I$. At the second order the Hamiltonian reads:
\begin{eqnarray}
H=H_0+\alpha(\omega ^2+\delta u^2_{1y}+\delta u^2_{2y}) + \beta (\omega ^2+\delta w^2_{1y}+ \nonumber \\
 \delta w^2_{2y}) + 2\gamma [\delta u_{1y}\delta u_{2y}+\delta w_{1y}\delta w_{2y}]   
\label{eq:F14}
\end{eqnarray}
The transition from phase $I$ to the multiferroic phase $II$ associated with the wave vector $\vec{k}_c$ occurs when $\frac{\partial H_0}{\partial k_c} = 0$ and $\mid \alpha (\vec{k}_c )\mid = \mid \gamma \mid $ or $\mid \beta(\vec{k}_c )\mid = \mid \gamma \mid $. These equations give, respectively, the equilibrium value of $\vec{k}_c$ in phase $I$ and the stability limits of phase $II$. In order to show that these limits exist and that a polar phase can actually be stabilized below phase $I$, we consider the situation where only interactions between first-neighbouring cells are involved. In this case the non-zero coefficients in Eq. (\ref{eq:F10}) are $h^{1-1}_0,h^{\pm 1\pm 1}_{\pm \vec{a}},h^{\pm \pm}_{\pm \vec{b}},h^{\pm \pm}_{\pm \vec{c}} $ and one has:
\begin{eqnarray}
\alpha (\vec{k}_c ) = (2h^{11}_{\vec{a}}+h^{1-1}_{\vec{a}}+h^{-11}_{\vec{a}})cos\vec{k}_c\cdot \vec{a}\nonumber \\
+(\vec{a}\rightarrow  \vec{b})+(\vec{a}\rightarrow  \vec{c}) \nonumber \\
\beta (\vec{k}_c ) = (2h^{11}_{\vec{a}}-h^{1-1}_{\vec{a}}-h^{-11}_{\vec{a}})cos\vec{k}_c\cdot \vec{a}\nonumber \\
+(\vec{a}\rightarrow  \vec{b})+(\vec{a}\rightarrow  \vec{c}) \nonumber \\   
\gamma  (\vec{k}_c ) = 4(h^{1-1}_{\vec{a}}-h^{-11}_{\vec{a}})sin \vec{k}_c\cdot \vec{a}\nonumber \\
+(\vec{a}\rightarrow  \vec{b})+(\vec{a}\rightarrow  \vec{c})
\label{eq:F15}
\end{eqnarray}
If $h^{11}_{\vec{a}}>0$ and $h^{11}_{\vec{b},\vec{c} }<0$ $H_0$ is minimum at $\vec{k}_c=(\frac{\pi }{a},0,0) $ and $\alpha $ becomes $h^{1-1}_0-(2h^{11}_{\vec{a}}+h^{1-1}_{\vec{a}}+h^{-11}_{\vec{a}})+(2h^{11}_{\vec{b}}+h^{1-1}_{\vec{b}}+h^{-11}_{\vec{b}})+(2h^{11}_{\vec{c}}+h^{1-1}_{\vec{c}}+h^{-11}_{\vec{c}})$, with  an analogous expression for $\beta $ and $\gamma $. When the $h^{1-1}_{\vec{a},\vec{b},\vec{c}}$ and $h_0$ coefficients vary, $\alpha \pm \gamma $ and $\beta \pm \gamma $ take positive \textit{and} negative values. Thus \textit{the non-polar phase $I$ and polar phase $II$ can both be stabilized by isotropic  exchange interactions}. Considering interactions between spins belonging to more distant unit-cells would stabilize commensurate or incommensurate, polar or non-polar phases depending on the values of the $h^{m,m'}_{\vec{n}-\vec{n}'}\delta_{\mu,\mu'}$ coefficients of the exchange Hamiltonian. However, since no assumption is made on the actual values of the interaction coefficients, one cannot exclude that in $h^{\pm 1\pm 1}_{\vec{a},\vec{b},\vec{c}}$ space, the region of the phase diagram in which the polar phase is stable would actually be forbidden. \\
\indent In summary, the specific features of multiferroic phases induced by the coupling of replicate order-parameters have been illustrated by the phenomenological description of the sequence of triclinic phases observed in FeVO$_4$.  The replication mechanism yields the onset of an additional lower symmetry ferroelectric phase in the phase diagram, the dephasing between the order-parameter replicates varying critically at the ferroelectric transition. Considering a single sublattice of FeVO$_4$ the stabilization of the ferroelectric phase has been shown to result from purely isotropic exchange interactions. It would be of interest to verify if this property is specific to mutiferroic systems involving a replication mechanism.\\
\indent The possibility of stabilizing additional phases by coupling two dephased order-parameters having the same symmetry, i.e., transforming as the same irreducible representation of the parent phase, was first noted in superconductors~\cite{ref:13} which involve a breaking of the continuous gauge symmetry. Another example of replication mechanism was described in ferroelectric liquid crystals~\cite{ref:14} in which the continuous rotational symmetry is broken at the transition to the ferroelectric phase. Although structural and magnetic phase transitions correspond to discrete symmetry breaking processes which do not give rise to additional phases by order-parameter replication, transitions to \textit{incommensurate} structural or magnetic structures permit a realization of the replication mechanism, as shown in the present work for the multiferroic transition in FeVO$_4$. This is because the thermodynamic features of the transition are driven by an \textit{effective} continuous group generated by the images of the crystallographic translations in the order-parameter space: The discrete translation group broken at the transition to the incommensurate structure has a \textit{dense} image in the order-parameter space acting as the continuous rotation group of the phason. In other words, for incommensurate wave-vectors the infinite discrete group of translations has the same effect on the form of the transition free-energy as a continuous group.\\
\\
\indent The authors are grateful to Laurent Chapon and Dmitry Khalyavin for helpful discussions

\end{document}